\documentstyle[multicol,epsf,pre,aps]{revtex}

\newcommand{\be}{\begin{eqnarray}}
\newcommand{\ee}{\end{eqnarray}}
\newtheorem{thm}{Case} 

\begin{document}
\date{\today}

\draft

\title{Stress condensation in crushed elastic manifolds}

\author{Eric M. Kramer\cite{byline} and Thomas A. Witten}
\address
{The James Franck Institute and the Department of Physics\\
The University of Chicago,Chicago, Illinois 60637}
\maketitle

\begin{abstract}

We discuss an $M$-dimensional phantom elastic 
manifold of linear size $L$ crushed into a small 
sphere of radius $R \ll L$ in $N$-dimensional space.   
We investigate the low elastic energy states 
of 2-sheets ($M=2$) and 3-sheets ($M=3$)
using analytic methods and lattice simulations.
When $N \geq 2M$ the  curvature energy is uniformly
distributed in the sheet and the strain energy is 
negligible. But when $N=M+1$ and $M>1$, both energies 
appear to be {\it condensed} into a network
of narrow $M-1$ dimensional ridges. The ridges
appear straight over distances comparable 
to the confining radius $R$.
\end{abstract}


\begin{multicols}{2}

It has long been known that a thin
elastic plate will develop narrow ridges
under a variety of compressive boundary conditions.
These ridges may be seen daily in the 
way a pillowcase or trouser leg deforms to wrap
its contents, often exhibiting a diamond
pattern familiar from compression studies of
thin metal cylinders \cite{Cyl}. The linear scars 
in a crumpled sheet of paper are also
a record of this mechanism \cite{Crumpl}.
Recently it was discovered that the structure
of these ridges  
could be accounted for using linear elasticity theory,
valid in the limit that the ridge length $X$
is much greater than the plate thickness $h$ \cite{WitLi,Lobkov}.
Witten and Li used a scaling argument
to predict that the ridge width $w \simeq h^{1/3}X^{2/3}$
and the total elastic energy $E\simeq Yh^{3}(X/h)^{1/3}$
where $Y$ is the Young's modulus \cite{WitLi}.
Lobkovsky {\it et al.} verified these scaling laws 
using both numerical simulations and an asymptotic
analysis of the Von Karman equations for a thin plate \cite{Lobkov}. 
His simulations showed that the 
material strains and curvatures decay rapidly to zero 
in the direction transverse to the ridge.
The length scale of this decay is the ridge 
width $w$, which goes to zero with
the thickness of the plate.  However, these ridges were analyzed
only in idealized, symmetrical deformations. Their applicability to
stochastically crumpled sheets has not been explicitly shown.
 
Ridge formation is a mode of spontaneous 
condensation of energy into a small subset
of the available volume.  As such it resembles
the spontaneous organization of
dislocations into grain boundaries in a strained 
crystal or the formation of Prandtl boundary 
layers in laminar fluid flow \cite{Dis,Hydro}. 

To understand the necessary conditions
for this condensation, and its consequences, it is useful 
to consider the general problem of an 
$M$-dimensional elastic manifold crushed by a 
hypersphere in $N$-dimensional space.
In Ref. \cite{Kram2} we derived the elastic 
energy functional of the manifold by 
considering the small thickness limit of 
an $N$-dimensional elastic solid with an extent $O(L)$
in $M$ directions and a thickness 
$h$ in $N-M$ transverse directions.
Using this formalism we found that a deformed 
hypersurface ($M=N-1$) under a specific 
boundary condition exhibits a
ridge with scaling properties analogous to those of 
Ref. \cite{Lobkov}.
In this Letter we explore the more general situation
of an $M$-sheet with a free boundary confined by a small
sphere. Using analytic and numerical methods
we demonstrate two complementary behaviors:
\begin{thm}[$N \geq 2M$]
The energy   $E_{c}$ associated with curvature is distributed {\it 
uniformly} in the manifold and the  energy $E_{s}$ associated with  
strain is negligible by comparison.
\end{thm}
\begin{thm}[$N=M+1>2$]
The  strain  energy and curvature energy are comparable, with
$E_{s}/E_{b}\simeq 0.2$. Both energies are  {\it condensed} into
a network of narrow,
$M-1$ dimensional ridges as described in  
Refs. \cite{Lobkov} and \cite{Kram2}. 
\end{thm}
In the remaining cases, namely $M+1 < N < 2M$, 
we anticipate that the  strain energy 
will also be comparable to the curvature energy 
\cite{Proof}. We have not investigated these cases
in detail.

The differential geometry of the deformed manifold
is conveniently discussed using a Euclidean coordinate patch 
on the {\it flat} manifold $(x_{\alpha}| \alpha \in [1,M])$,
called the manifold coordinate system \cite{DiffGeom}.
Any deformation may then be written as a map $\vec{r}(x)$
from the manifold coordinates to $N$-dimensional Euclidian 
space.  
The strain tensor is 
$u_{\alpha\beta}(x)=(1/2)
(\partial_{\alpha}\vec{r}\cdot\partial_{\beta}\vec{r}
-\delta_{\alpha\beta})$ and the extrinsic curvature 
tensor is
$\vec{K}_{\alpha\beta}=\partial_{\alpha}\partial_{\beta}\vec{r}$
\cite{DiffGeom}. Under the usual assumptions of
linear elasticity theory
($u_{\alpha\beta} \ll 1$ and 
$\partial_{\gamma}u_{\alpha\beta}\ll |\vec{K}_{\mu\nu}| \ll 1/h$
\cite{LL})
the elastic energy functional becomes
\be
\begin{array}{rl}
E=\int dx^{M} & \left\{
 \mu \left( (u_{\alpha\beta})^{2} 
+ c_{0} (u_{\alpha\alpha})^{2} \right)  \right.  \\
& \left. + \kappa \left( \vec{K}_{\alpha\beta}
\cdot\vec{K}_{\alpha\beta}
+ c_{0} \vec{K}_{\alpha\alpha}\cdot\vec{K}_{\beta\beta} \right) 
\right\}
\end{array}
\label{eq:etot}
\ee
where  $\mu \simeq Yh^{N-M}$ is an effective Lam\'{e}  coefficient,
 $\kappa \simeq Yh^{N-M+2}$ is an effective bending rigidity,
and $c_{0}$ is a dimensionless constant \cite{Kram2,units}. 
Summation over repeated indices is implied.
We shall refer to the term quadratic in the strains as
the  strain energy $E_{s}$, and the term
quadratic in the curvatures as the curvature 
energy $E_{c}$. We consider only {\it phantom}
(not self-avoiding) manifolds in this paper, since
we don't expect self-avoidance to alter the qualitative
conclusions. Our compressive boundary condition is a frictionless, 
hard-walled sphere with an initial
radius $R_{i} >L$. To crush the manifold the radius 
is slowly decreased to a final value $R \ll L$.

In our simulations
we model a thin elastic 2-sheet ($M=2$) using 
a triangular network of nodes connected by springs (see Fig. 1),
following the work of Seung and Nelson \cite{Seung} . 
The discretized  strain 
energy is the sum over 
the Hooke's law energy of each spring 
\be
E_{s}=\sum_{i} \frac{1}{2}c_{s}(l_{i}-l_{0})^{2}
\ee
where $c_{s}$ is a  spring
constant controlling the strain energy
$l_{i}$ is the length of spring $i$, and $l_{0}$ is the lattice constant.

\begin{figure}
\centerline{\epsfxsize=8.5cm \epsfbox{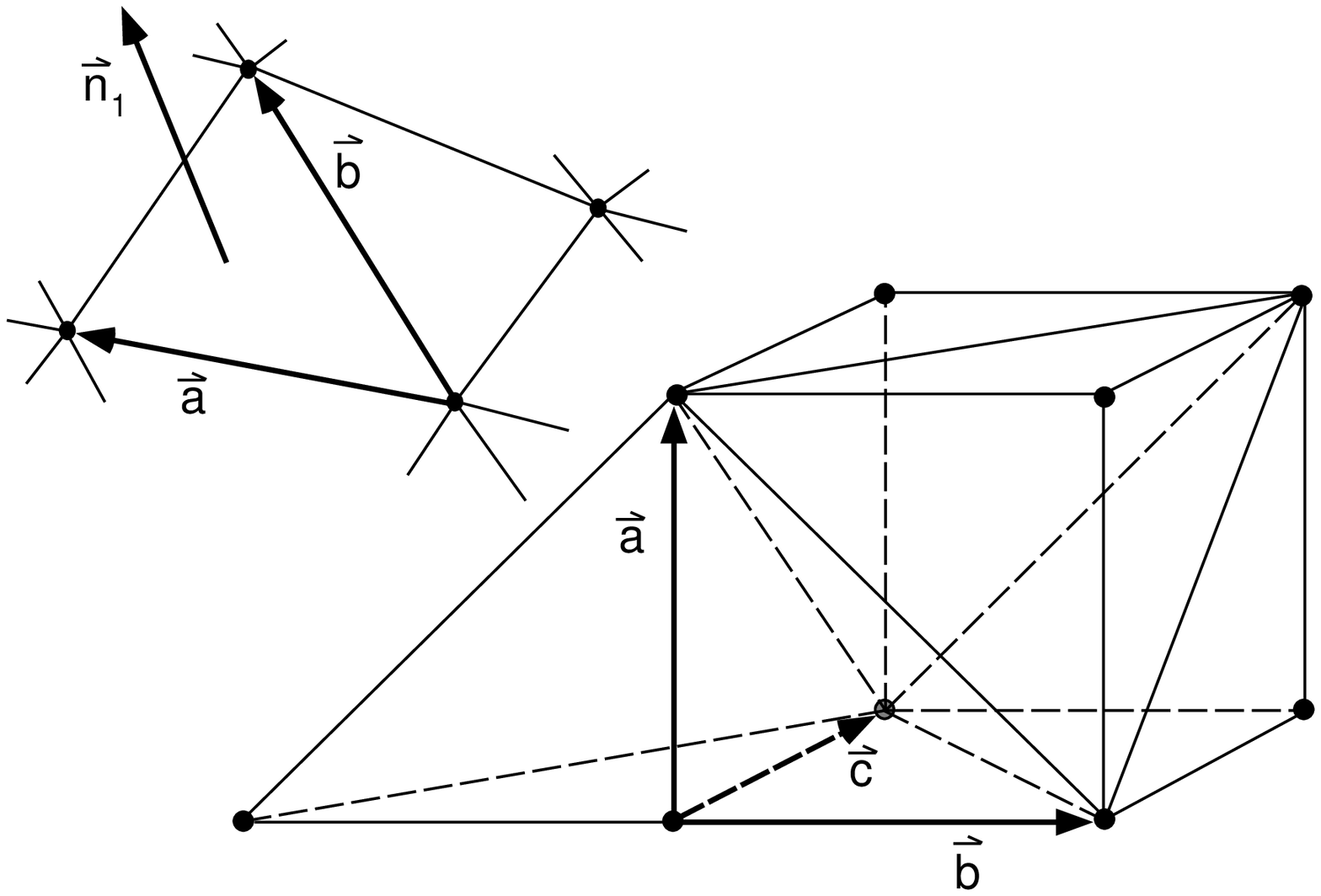}}
{FIG. 1. A portion of the triangular lattice
and the simple cubic lattice used to discretize 
a 2-sheet ($M=2$) and a 3-sheet ($M=3$) respectively.}
\end{figure} 

To discretize the  curvature 
energy it is convenient to
begin by defining the normal tensor for a  two-dimensional lattice in 
 $N$-dimensional Euclidean space.
 If $\vec a(x)$ and $\vec b(x)$ are two independent vectors tangent to the
2-sheet at $x$, then we may define a normal tensor via
\be 
({\sf n}[a,b])_{i_{1} \cdots i_{N-2}}=
\epsilon_{i_{1} \cdots i_{N}}a_{i_{N-1}}
b_{i_{N}} ,
\ee
where  $a_i$ and $b_i$ are the Cartesian components of $\vec a$ and $\vec b$ and $\epsilon$ is the antisymmetric Levi-Civita 
tensor. In 3-space this reduces to the usual
normal vector ${\sf n}=\vec{a}\times\vec{b}$.
Defining the inner product 
\be
{\sf n}_{1}\cdot {\sf n}_{2} = \frac{1}{(N-M)!}
({\sf n}_{1})_{i_{1} \cdots i_{N-M}}
({\sf n}_{2})_{i_{1} \cdots i_{N-M}}
\label{eq:InProd}
\ee
with $M=2$, the unit normal tensor  may be expressed
$\hat{\sf n}={\sf n}/\sqrt{\sf n \cdot \sf n}$. By substituting
$\vec{a}=\partial_{1}\vec{r}$ and $\vec{b}=\partial_{2}\vec{r}$,
it is straightforward to prove that 
$\partial_{\alpha}{\sf n}\cdot\partial_{\alpha}{\sf n}
=\vec{K}_{\alpha\beta}\cdot\vec{K}_{\alpha\beta}$.
The  curvature 
energy  can therefore be represented as
the square of the discrete derivative
\be
E_{c}=\sum_{\langle ij\rangle} \frac{1}{2}c_{b}
|\hat{\sf n}_{i}-\hat{\sf n}_{j}|^{2} 
\label{eq:ebsim}
\ee
where $c_{b}$ is a bending constant, $\hat{\sf n}_{i}$ 
is the unit normal of triangle $i$ (see FIG. 1), 
and the sum is taken over each
pair of triangles which share a common edge.
Comparison with Eq. (\ref{eq:etot}) gives $\mu \simeq c_{s}/l_{0}^{2}$,
$\kappa \simeq c_{b}/l_{0}^{2}$, and $c_{0}=0$
\cite{Seung}. The effective thickness of the  2-sheet
is determined by the ratio
$c_{b}/c_{s} \simeq h^{2}$. We take 
$(c_{b}/c_{s})^{1/2} \gtrsim l_{0}/6$ to minimize 
lattice effects in our simulations.

The discrete model of an elastic 3-sheet ($M=3$) is qualitatively 
similar. The  3-sheet is a simple cubic lattice of nodes 
connected by springs. As shown in Fig. 1, there are 
edge springs with constant $c_{e}$ and length $l_{0}$ 
and diagonal springs with constant $c_{d}$ and 
length $\sqrt{2}l_{0}$. The requirement that the  lattice
be elastically isotropic for small strains
is $c_{d}=2 c_{e}$. To calculate the  curvature 
energy
we first define the normal tensor on a 3-manifold 
in  $N$-space
\be 
({\sf n}[a,b,c])_{i_{1} \cdots i_{N-3}}=
\epsilon_{i_{1} \cdots i_{N}}a_{i_{N-2}}
b_{i_{N-1}}c_{i_{N}}
\ee
where $\vec{a}(x)$, $\vec{b}(x)$, and $\vec{c}(x)$ span
the tangent space of the manifold at $(x)$.
Using the inner product Eq. (\ref{eq:InProd})
with $M=3$, we again have 
$\partial_{\alpha}{\sf n}\cdot\partial_{\alpha}{\sf n}
=\vec{K}_{\alpha\beta}\cdot\vec{K}_{\alpha\beta}$.
Each unit cube is conceived as five adjacent 
tetrahedra with the springs
as edges. Each cube then has four ``corner'' tetrahedra
surrounding one ``middle'' tetrahedron.
The sum in Eq. (\ref{eq:ebsim}) is taken
over all pairs of tetrahedra that share a common triangular 
face (see Fig. 1). Lastly, we relate the bending 
constant $c_{cc}$ for a pair
of corner tetrahedra to the bending constant $c_{cm}$
for a corner-middle pair. Isotropy requires $c_{cm}=2 c_{cc}$.

The initial condition in our simulations is  
a hexagonal  2-sheet ($M=2$) or a roughly spherical 3-sheet  
($M=3$) with a small random 
displacement added to the positions of the nodes \cite{fluctuations}.
We model a hyperspherical container of radius 
$r_{0}$ with the potential 
$V_{sphere}=\sum_{i} (r_{i}/r_{0})^{12}$, where
$r_{i}$ is the distance from the origin to node $i$.
We use the rms radius of the manifold as our measure of the confining radius 
$R$: $R^2 = \sum_i r_i^{\,2}~/~(\sum_i1)$.
To crush the manifold, the
value of $r_{0}$ is repeatedly decreased, and the 
energy of the manifold minimized at each step using a conjugate
gradient routine \cite{NumRec}. Using this method a hexagonal 
 2-sheet with long diameter $L=160 l_{0}$ may be crushed 
to a radius $r_{0}=20 l_{0}$ in  
three days of CPU time on an IBM RISC 6000.
An approximately spherical  ball with diameter $L = 30 l_{0}$
may be crushed to $r_{0}=6.0 l_{0}$ in about five days.

The simplest instance of Case 1 is a thin
elastic rod crushed within a circle in 2-space.
As the radius of the circle is decreased the rod buckles and develops a 
curvature of $O(1/R)$. Because the rod is free to reptate parallel to its 
length, its  strain energy remains identically zero.  Analogously, a thin 
plate in  4-space can curl without  strain into an arbitrarily small 
3-sphere. One possible embedding which demonstrates this is   
$\vec{r} = \rho [\cos (x_{1}/\rho), \sin (x_{1}/\rho),
\cos (x_{2}/\rho), \sin (x_{2}/\rho) ]$ where 
$\rho = R/\sqrt{2}$. Note that $r^{2}=R^{2}$ everywhere.
One can verify by substitution 
that the strain tensor  $u_{\alpha \beta}$
is identically zero, 
$|\vec{K}_{11}(x)|=|\vec{K}_{22}(x)|=\sqrt{2}/R$, and
$E_{tot} \simeq \kappa (L/R)^{2}$. The key feature 
of this embedding is the separation of the manifold
coordinates into independent, 2-dimensional subspaces
of  4-space. Analogously, whenever $N \geq 2M$ we may write
the embedding  
\be
\vec{r} =  \rho [ & \cos (x_{1}/\rho), \sin (x_{1}/\rho),
\cos (x_{2}/\rho), \sin (x_{2}/\rho), 
\nonumber \\
& \ldots,\cos (x_{M}/\rho), \sin (x_{M}/\rho),0,0,\ldots ]
\label{eq:torus}
\ee
where $\rho = R/\sqrt{M}$. 
This embedding is an existence proof that
the manifold need not strain during the
compression. It is also the {\it global} minimum
of the elastic energy at fixed $R$ (neglecting corrections
within $R$ of the manifold boundary). 
This deformation removes the isotropy of the manifold 
whenever $M>1$.  Thus, {\it e. g.} in order to return 
to the starting position $\vec r$ one must move
$\sqrt{2}$ as far in the $(1,1)$ direction as 
in the $(1,0)$ direction. With
this broken symmetry comes a degeneracy.  
For every distinct orientation of the 
coordinate system $\{x_1, ..., x_M\}$, 
Eq. (\ref{eq:torus}) yields a new
minimum-energy configuration.

Our simulations of a  2-sheet in 4-space 
and a 3-sheet in 6-space behave 
as anticipated for Case 1. 
In both instances we observe the minimum energy embedding 
discussed above, with a  curvature 
energy density uniform to 10\% and 
negligible  strain 
energy $E_{s}\lesssim 10^{-3} E_c$.
The only significant deviation from the ideal
embedding is that the manifold flattens out 
within $R$ of the manifold edge.
As a result, the  curvature 
energy density decays
to zero near the edge and there 
is a systematic downward correction to the total energy
of the form $E_{tot} \sim (L-(\mbox{const.})R)^{M}/R^{2}$.
Depending upon the initial condition,
we also observe several metastable 
states with a nonuniform curvature 
energy density and 
a higher total energy. As compression proceeds
these metastable states make the transition to the 
minimum energy embedding. 
 
Our understanding of Case 2 deformations is incomplete. 
It is based on previous analytic and numerical 
work for a variety of simple deformations, 
including some on general hypersurfaces \cite{Lobkov,Kram2}.  
These cases all show the condensation 
of the elastic energy 
into a fraction $O(h/X)^{1/3}$ of the manifold volume,
suggesting that a generic compression 
yields similar condensation. 
The simple cases also exhibit a strict proportionality 
between strain and curvature energy: 
$\lim_{ (h/X) \rightarrow 0} E_s = E_{c}/5$.

We simulated Case 2 deformations using a 2-sheet in 3-space and
a 3-sheet in  4-space. For the 2-sheet we see 
the  anticipated spontaneous formation of a network
of linear ridges. Fig. 2 shows the  curvature 
energy density in the manifold
coordinates of a hexagonal  2-sheet
with long diameter $L=160 l_{0}$ crushed to a radius 
$R = 27 l_{0}$.
About 40\% of the total  curvature energy
occupies just 2\% of the total area. This is due
to the formation of point vertices in the  2-sheet.
These vertices are the tips of cone-like deformations.
The next 40\% occupies 20\% of the area. We see that 
this energy is  condensed into a network of narrow ridges
which connect the vertices. The  strain 
energy density
is similarly localized. The ratio $E_s/E_c \approx 10$\%.
This is consistent with the deviations from
the asymptotic value 
$E_s/E_c =1/5$ reported by Lobkovsky
for short ridges $X \approx 10 h$ \cite{Lobkov}. 

\begin{figure}
\centerline{\epsfxsize=9cm \epsfbox{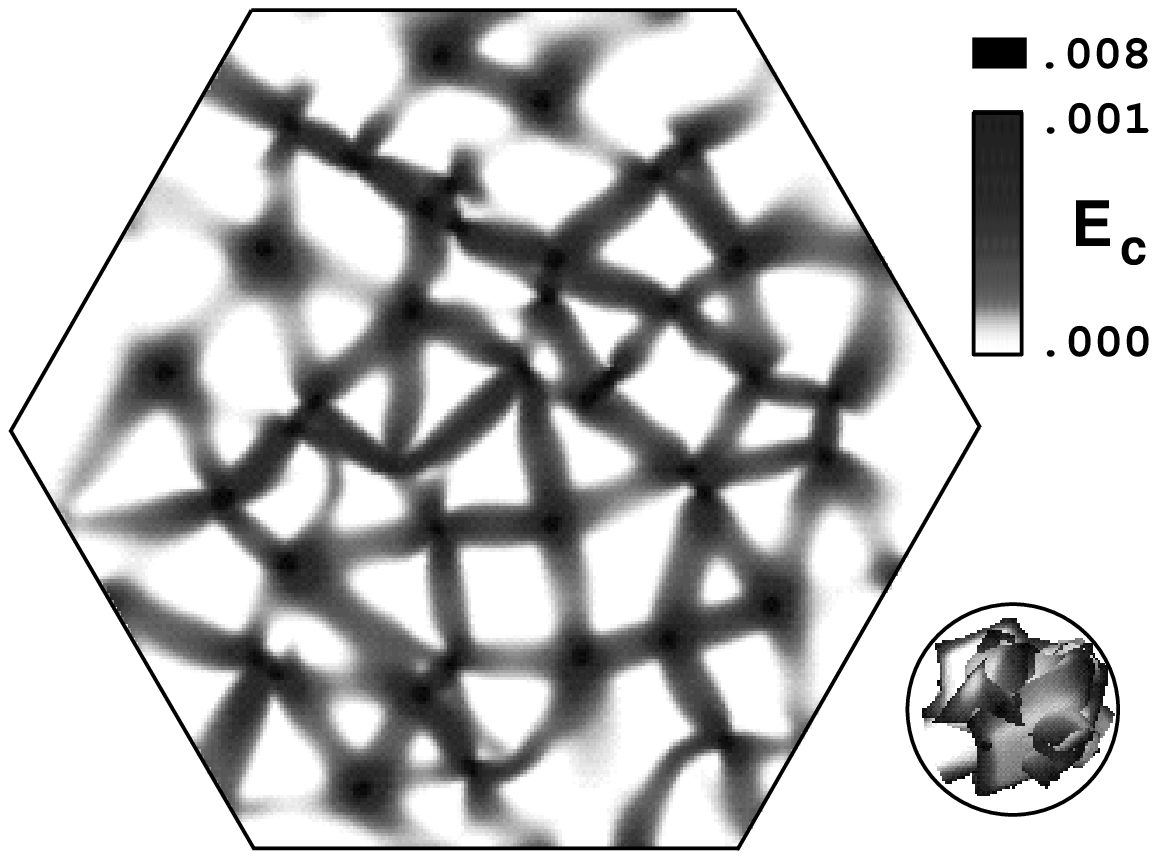}}
{FIG. 2. Curvature energy distribution in a hexagonal
2-sheet with moduli $c_{s}=1.0$ and $c_{b}=.025$
and long diameter $L=160 l_{0}$ crushed 
to a radius $r_{0}=27 l_{0}$. Darker regions have higher energy density.  
The crumpled configuration is shown in the lower right.}
\end{figure} 

Fig. 2 suggests that the ridges forming the network are of comparable 
length. This length is similar to the diameter of the confining sphere 
shown in the lower right corner. 
If we make the scaling hypothesis that the 
mean ridge length $\bar{X} \simeq R$, then 
the number of ridges in the  2-sheet scales like $(L/R)^{2}$. 
Since the energy of a single ridge scales as 
$Yh^{3}(R/h)^{1/3}$ \cite{Lobkov},
the total elastic energy $E_{tot}$ should obey
$E_{tot} \simeq YhL^{2}(h/R)^{5/3}$.
Our simulation results are consistent with 
$E_{tot} \sim R^{-5/3}$ \cite{uncertainty}.  
However, due to boundary effects
and the small values of the aspect ratio
$L/h \lesssim 300$
the evidence for this scaling is only suggestive.

Our simulations of an elastic  3-sheet in 4-space
attained aspect ratios
of only $L/h \simeq 30$.
Since the ridge properties derived in Ref. \cite{Kram2} 
are well-defined only in the limit that the ridge 
is much longer than $h$, these simulations 
provide only qualitative support for Case 2.
The simulations show the
anticipated concentration of strains and curvatures 
into linear and planar structures.
Fig. 3 shows two  curvature energy isosurfaces 
in the manifold coordinates of a  3-sheet 
with diameter $L=30 l_{0}$ crushed to 
$r_{0} = 12 l_{0}$. The upper isosurface 
encloses 65\% of the total  curvature 
energy and 16\% of the volume. We see distinct
planar structures. The lower isosurface 
encloses 23\% of the curvature energy and 
only 3\% of the volume. The energy is concentrated
on the set of lines where planes intersect.
The lines are the analogs of the point-like 
vertices in FIG. 2. The ratio $E_s/E_c$ is 15\%.

We anticipate scaling behavior for the energy of 
a crushed hypersurface analogous to that for a crushed 2-sheet.  
If the linear size of the ridges is roughly the 
diameter of the confining sphere, we may 
generalize the argument for the 2-sheet to infer 
$E_{tot} \simeq Y h L^{M}(h/R)^{5/3}$ \cite{Kram2}. 
The limitations of our simulation prevented us 
from testing this prediction.

\begin{figure}
\centerline{\epsfxsize=6.7cm \epsfbox{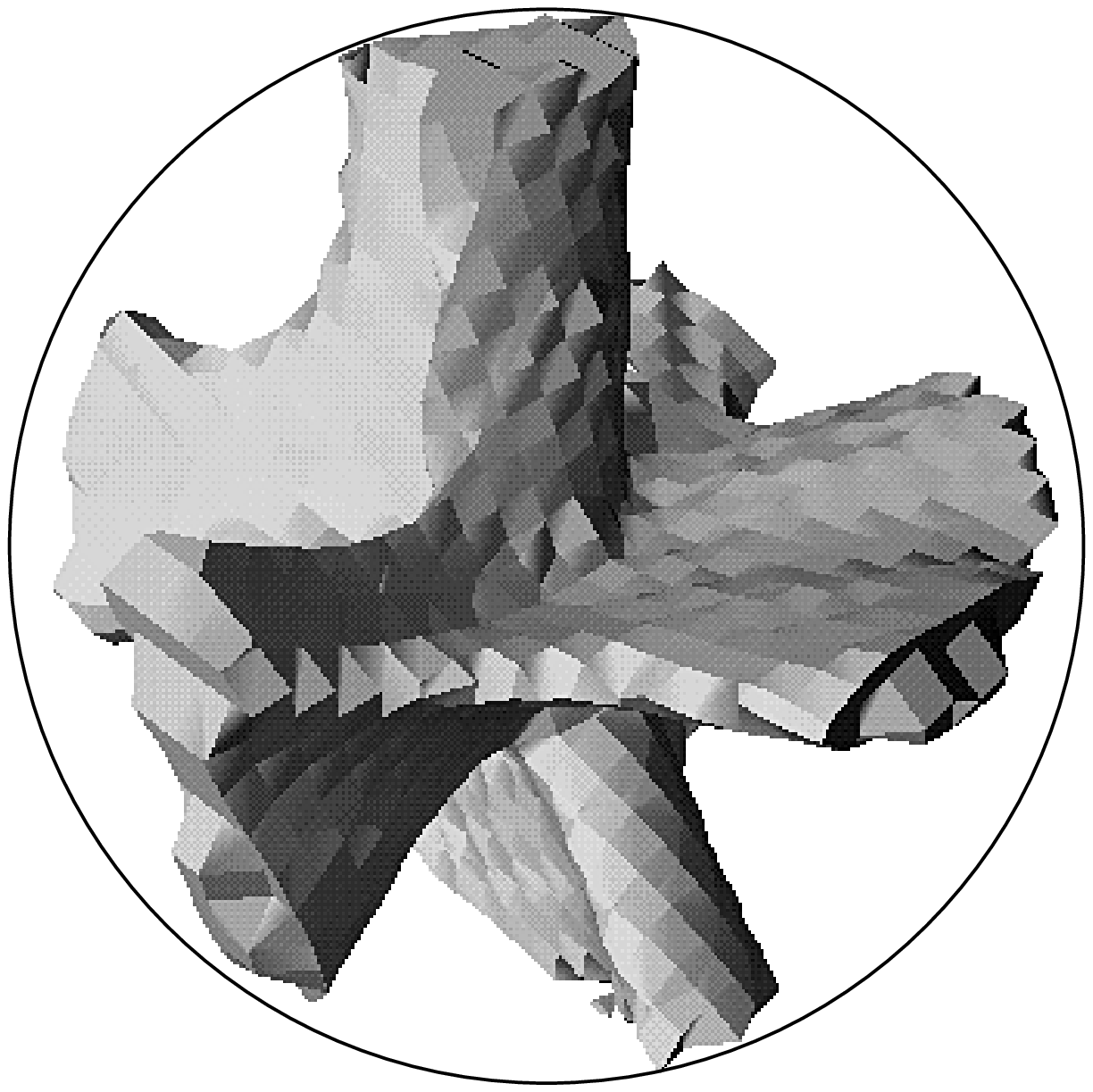}}
\centerline{\epsfxsize=6.7cm \epsfbox{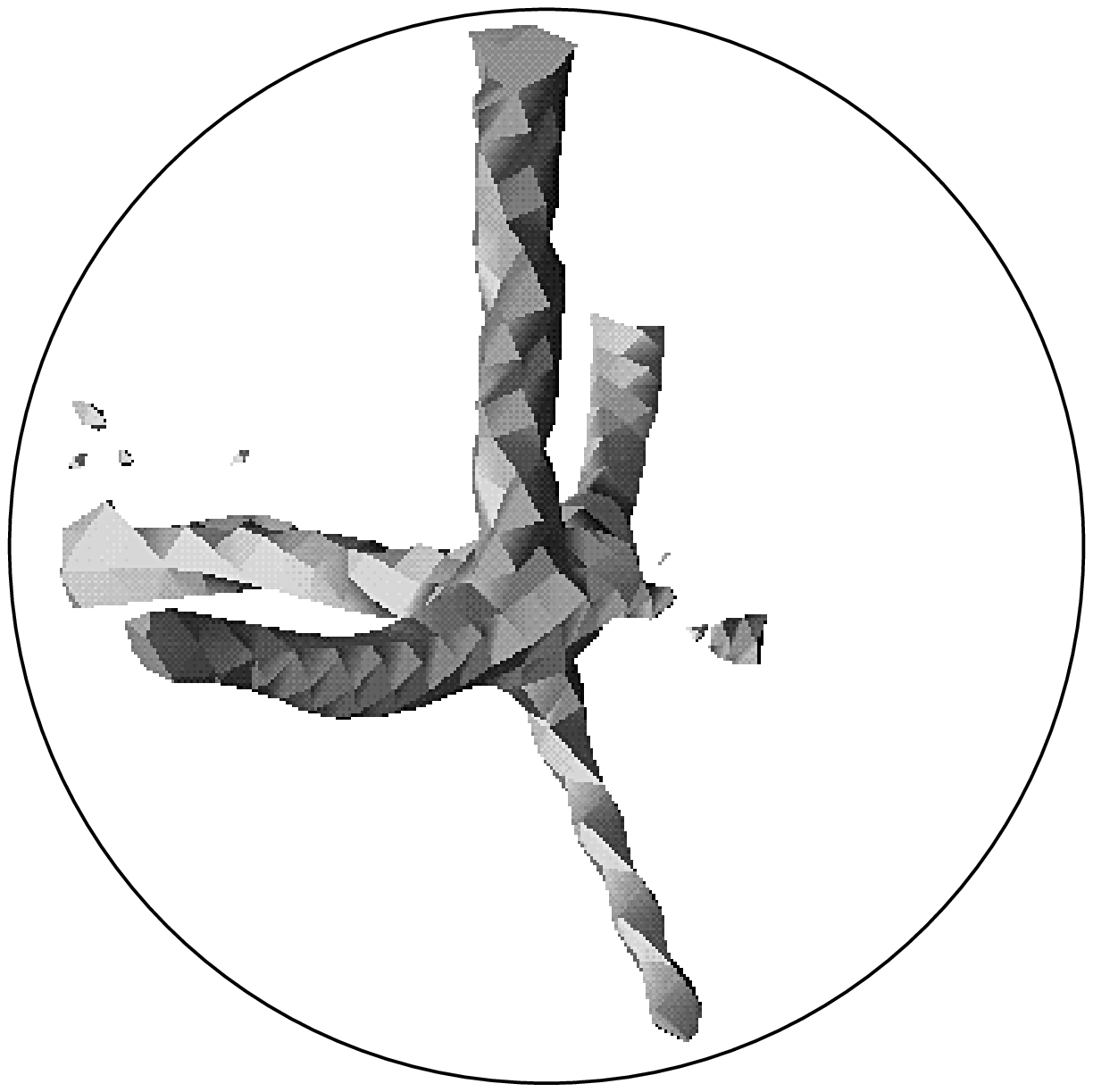}}
{FIG. 3. Curvature energy isosurfaces in the manifold 
coordinates of an elastic 3-sheet  ($L=30 l_{0}$) in 4-space crushed
to a radius $r_{0}= 13 l_{0}$. The top (bottom) surface
encloses 65\% (23\%) of the total energy in
16\% (3\%) of the volume.}
\end{figure} 

This work has implications for real crumpled sheets.  
It is the first exploration of the  
distribution of energy in these sheets and 
it offers qualitative support for the model of
a crumpled sheet as a network of ridges
\cite{Lobkov}. The possibility of crushing without 
strain in high dimensions may be relevant for studies 
of thermal crumpling in general
dimensions \cite{Kantor}.
More broadly, this work identifies a new 
mechanism of energy condensation
into a small subspace of an available volume.  
This condensation happens in arbitrary spatial 
dimensions in one of the simplest organizations
of matter -- an elastic manifold.  Increasingly, the behavior
of simple manifolds in general spatial dimensions is 
invoked to account for fundamental processes \cite{Ed.Witten}.  
The symmetry-breaking and ridge-forming 
mechanisms explored here may prove
relevant for understanding such behavior.

\acknowledgments
The authors thank
Robert Geroch, Alex Lobkovsky, and Jeff Harvey 
for helpful discussions. 
This work was supported in part
by the NSF through Grants No. DMR-9400379 and
DMR-9528957.

\end{multicols}


\begin{references}
\bibitem{Proof} The proof of this conjecture for 
$N=M+1$ and $N=M+2$ will be presented in a future paper.
\bibitem{units} The Young's modulus has units of energy
per unit $N$-volume, so $\mu$ has units of energy per unit
$M$-volume.
\bibitem{uncertainty} The simulation results are 
also consistent with $E_{tot}\sim R^{-2}$, 
the dependence expected for Case 1.
\end{references}
\end{document}